\documentclass[conference]{IEEEtran}
\IEEEoverridecommandlockouts
\usepackage{cite}
\usepackage{amsmath,amssymb,amsfonts}
\usepackage{amsmath}
\usepackage{balance}
\usepackage{algorithmic}
\usepackage{graphicx}
\usepackage{textcomp}
\usepackage{xcolor}
\def\BibTeX{{\rm B\kern-.05em{\sc i\kern-.025em b}\kern-.08em
    T\kern-.1667em\lower.7ex\hbox{E}\kern-.125emX}}
\begin{document}

\title{Grid-Forming Control of Power Converters: Equivalence Proof through Simplified Models}

\author{\IEEEauthorblockN{Aidar Zhetessov}
	\IEEEauthorblockA{\text{Dept. of Electrical and Computer Engineering} \\
		\text{1415 Engineering Drive}\\
		Madison, WI 53706}}

\maketitle

\begin{abstract}
This work establishes the equivalence of selected grid-forming control algorithms within the context of simplified theoretical models. Considered algorithms are droop control, Virtual Synchronous Machine (VSM) and matching control. It is shown that nodal and network dynamics under those regulators boil down to the same equations near the selected (trivial) nominal operating point. Finally, some practical insights on each regulator dynamics and an outlook are provided.
\end{abstract}


\section{Introduction}
Sustainable development pushes the technology towards grid integration of renewable energy sources (RES) \cite{Kroposki}. With more RES penetrating the grid, some form of grid formation/support is required from the RES interfacing converters for a stable grid operation \cite{Peng}. As a response to that necessity, several converter control algorithms have been developed with a general term “grid-forming” control serving as an umbrella definition for those \cite{Rocabert}, \cite{Poolla}. Naturally arising questions here are how do those grid-forming techniques compare and which one to use? Numerous detailed studies have addressed those (e.g. \cite{Tayyebi}). 

The small project at hand aims at contributing to the aforementioned works by theoretically establishing the equivalence between the grid-forming control techniques and gaining some practical insights on controller selection/tuning. Considered techniques are droop, Virtual Synchronous Machine (VSM) and matching control. To simplify the analysis, network AC power flows are linearized around the nominal (trivial) network operating point. Moreover, inner voltage-current regulators of the converters are assumed to be ideal.

The work is organized as follows: Section \ref{s:Node} introduces nodal dynamics with grid-forming regulators and establishes equivalence. Section \ref{s:Network} analyzes network dynamics and presents some insights on grid-forming regulator selection/tuning from a network perspective. Finally, Section \ref{s:Conclusion} summarizes the findings and shapes an outlook.

\section{Nodal Equivalence of Grid-Forming Controls} \label{s:Node}
This section presents the nodal dynamics of the power converters under three different grid-forming controllers: VSM, droop and matching. VSM is referred to as reference regulator. Unless otherwise stated, all node dynamics are in dq-frame (rotating at nominal grid frequency $\omega_0$), not linearized (full) and virtual (state integration happens numerically).

\subsection{Virtual Synchronous Machine}
VSM nodal dynamics are summarized in \eqref{eq:VSM} for an arbitrary node $k$ of the network, represented as a graph:

\begin{equation}
\begin{split}
\frac{d}{dt} \theta_k &= \omega_k
\\
M_k \frac{d}{dt} \omega_k &= -D_k \omega_k + P_k^* - \tilde{P}_k
\\
\tilde{P}_k &\approx P_k
\\
\tau_f \frac{d}{dt} \tilde{Q}_k &= Q_k - \tilde{Q}_k
\\
V_{m,k} &= V_{m,k}^* + R_q \big(Q_k^* - \tilde{Q}_k\big)
\end{split}
\label{eq:VSM}
\end{equation}

Here the dynamics are given as they are implemented in numerical integration. In particular, five variables ($\omega_k, \theta_k, V_{m,k}, Q_k, P_k$) are manipulated and three ultimate state variables $\omega_k, \theta_k, V_{m,k}$ are passed further to the inner voltage/current loops for subsequent physical implementation on the network node. Power variables $Q_k$ and $P_k$ signify physical reactive/active powers entering/leaving the node, while their counterparts $\tilde{Q}_k$ and $\tilde{P}_k$ represent their averaged (low-pass filtered) measurements. $\tau_f$ parameter in $Q_k$ relation is the low-pass time-constant. Note that active power filter dynamics are neglected because the subsequent virtual inertia dynamics usually are much longer in time ($M_k/D_k >> \tau_f$) so that by the time power terms have any effect on $\omega_k$ dynamics, the active power is already stabilized.

By eliminating the measurement dynamics one can reach to the system representation of solely state variable dynamics \eqref{eq:VSMfinal}:

\begin{equation}
\begin{split}
\frac{d}{dt} \theta_k &= \omega_k \\
\tau_f \frac{d}{dt} V_{m,k} &= R_q \big(Q_k^* - Q_k\big) + \big(V_{m,k}^* - V_{m,k}\big) \\
M_k \frac{d}{dt} \omega_k &= -D_k \omega_k + P_k^* - P_k
\end{split}
\label{eq:VSMfinal}
\end{equation}

State dynamics of the form \eqref{eq:VSMfinal} can be referred to as reference nodal dynamics, through which the regulator equivalence is established.

\subsection{Droop Control}
Virtual nodal dynamics of the droop controller (as implemented numerically) are given in \eqref{eq:Droop}:

\begin{equation}
\begin{split}
\frac{d}{dt} \theta_k &= \omega_k
\\
\omega_k &= R_p \big(P_k^* - \tilde{P}_k\big)
\\
\tau_f \frac{d}{dt} \tilde{P}_k &= P_k - \tilde{P}_k
\\
\tau_f \frac{d}{dt} \tilde{Q}_k &= Q_k - \tilde{Q}_k
\\
V_{m,k} &= V_{m,k}^* + R_q \big(Q_k^* - \tilde{Q}_k\big)
\end{split}
\label{eq:Droop}
\end{equation}

Here, instead of mimicking the synchronous machine relations, dq-frame angular speed $\omega_k$ is made directly proportional to the estimated power mismatch between the reference power and the nodal measurement. Therefore, low-pass filter response on active power measurement can not be neglected. From \eqref{eq:Droop}, one arrives at the state variable dynamics by eliminating measurement variables:

\begin{equation}
\begin{split}
\frac{d}{dt} \theta_k &= \omega_k \\
\tau_f \frac{d}{dt} V_{m,k} &= R_q \big(Q_k^* - Q_k\big) + \big(V_{m,k}^* - V_{m,k}\big) \\
\underbrace{\frac{\tau_f}{R_p}}_{M_{dr,k}} \frac{d}{dt} \omega_k &= -\underbrace{\frac{1}{R_p}}_{D_{dr,k}} \omega_k + P_k^* - P_k
\end{split}
\label{eq:Droopfinal}
\end{equation}

From \eqref{eq:Droopfinal} it can be seen that droop state dynamics have the same form as in VSM case, thus proving equivalence. The corresponding droop inertia constant $M_{dr,k} = \tau_f/R_p$ is the ratio of the filter time-constant and the active power droop gain. The corresponding droop damping constant $D_{dr,k} = 1/R_p$ is the inverse of the active power droop gain. Similar results were obtained in \cite{DArco}.

With the filter time-constant assumption $\tau_f << M_k/D_k$ one can conclude that for droop control the nodal angle/frequency dynamics are much faster in converging to a steady-state compared to the VSM. On the other hand, one would expect $M_{dr,k} < M_k$ so that the initial rate of change of frequency (ROCOF) is greater in case of droop control, resulting in greater transient frequency overshoots.

\subsection{Matching Control}
Matching control is a relatively new alternative to the established counterparts presented above. Some work on matching was presented in \cite{Curi}, \cite{Arghir}. The virtual/physical nodal dynamics of matching control are presented in \eqref{eq:Matching}:

\begin{equation}
\begin{split}
\frac{d}{dt} \theta_k &= \omega_k \\
V_{DC,k} &= \tilde{V}_{DC,k} \\
\tau_f \frac{d}{dt} \tilde{Q}_k &= Q_k - \tilde{Q}_k \\
\omega_k &= K_{\theta} \big(\tilde{V}_{DC,k} - V_{DC,k}^*\big) \\
V_{m,k} &= V_{m,k}^* + R_q \big(Q_k^* - \tilde{Q}_k\big) \\
C_{DC,k} \frac{d}{dt} V_{DC,k} &= i_{DC,k} - \frac{1}{V_{DC,k}} P_k
\end{split}
\label{eq:Matching}
\end{equation}

Here, unlike techniques above, the frequency is directly related to the DC/AC converter DC voltage mismatch rather than to the active power imbalance in some form. Another differentiating feature is the fact that DC voltage dynamics (last equation in \eqref{eq:Matching}) is physical rather than virtual, meaning that the dynamics describe physical capacitor charing/discharging rather than its virtual counterpart. Alternatively, one could claim that the DC voltage integration forward in time occurs "physically" rather than numerically. As before, the remaining dynamics are still virtual, implying the need for DC voltage measurement. In \eqref{eq:Matching}, DC voltage measurement was approximated without low-pas filter as in case of active power in VSM dynamics. In this case, however, it is not due to subsequent dynamic time-scale separation, but due to the fact that physical DC voltage dynamics already model averaged voltage, neglecting the high-frequency switching ripple. AC voltage magnitude dynamics remain unchanged.

Presented DC-side voltage dynamics have a DC current injection $i_{DC,k}$ coming from the preceding converter/power source and serving as a control handle for DC voltage dynamics. Using this handle to establish a DC voltage proportional control results in equivalent state nodal dynamics as in VSM and droop as it will be shown in the following. DC voltage controller equations are as follows \eqref{eq:DCcontrol}:

\begin{equation}
\begin{split}
i_{DC,k} &= \tilde{i}_{DC,k}\\
\tilde{i}_{DC,k} &= i_{DC,k}^* + K_{DC} \big(V_{DC,k}^* - \tilde{V}_{DC,k}\big)
\end{split}
\label{eq:DCcontrol}
\end{equation}

Here $\tilde{i}_{DC,k}$ corresponds to the regulator output, while $i_{DC,k}$ is its physical manifestation. Solving together \eqref{eq:Matching}-\eqref{eq:DCcontrol} for state dynamics as before, one arrives at \eqref{eq:Matchingfinal}:

\begin{equation}
\begin{split}
\frac{d}{dt} \theta_k &= \omega_k \\
\tau_f \frac{d}{dt} V_{m,k} &= R_q \big(Q_k^* - Q_k\big) + \big(V_{m,k}^* - V_{m,k}\big) \\
\underbrace{\frac{C_{DC,k} V_{DC,k}^*}{K_{\theta}}}_{M_{DC,k}} \frac{d}{dt} \omega_k &= - \underbrace{\frac{K_{DC} V_{DC,k}^*}{K_{\theta}}}_{D_{DC,k}} \omega_k + ... \\
&... + \underbrace{V_{DC,k}^* i_{DC,k}^*}_{P_k^*} - \underbrace{\frac{V_{DC,k}^*}{V_{DC,k}}}_{\approx 1} P_k
\end{split}
\label{eq:Matchingfinal}
\end{equation}

Again, from \eqref{eq:Matchingfinal} one can observe equivalence to the VSM/droop state dynamics with $M_{DC,k}$ and $D_{DC,k}$ depending on DC-link physical parameters $C_{DC,k}, V_{DC,k}^*$ as well as control gains $K_{DC},K_{\theta}$.

Using the control gains one could shape the equivalent inertia and damping parameters to achieve the desired nodal transient performance. For instance, to limit the ROCOF one could reduce the $K_{\theta}$ gain, thus increasing the equivalent inertia $M_{DC,k}$. To reduce the steady state convergence time one could increase $K_{DC}$ gain. It would reduce the transient time-constant $\tau_{DC,k} = M_{DC,k}/D_{DC,k} = C_{DC,k}/K_{DC}$ and lead to faster convergence of nodal states. 

Next section provides more insights on desired nodal parameters and gains. As of now, the equivalent inertias and damping parameters of three grid-forming controllers are summarized in Table \ref{t:tab1}.

\begin{table}[htbp]
	\caption{Equivalent Inertia and Damping Parameters}
	\begin{center}
		\begin{tabular}{|c|c|c|c|}
			\hline
			\textbf{Parameter} &\multicolumn{3}{|c|}{\textbf{Control Technique}} \\
			\cline{2-4} 
			$\big(M_k , D_k\big)$ & \textbf{\textit{VSM}} & \textbf{\textit{Droop}} & \textbf{\textit{Matching}} \\
			\hline
			$M_k$ & $M_k$ & $\frac{\tau_f}{R_p}$ & $\frac{C_{DC,k}V_{DC,k}^*}{K_\theta}$ \\
			\hline
			$D_k$ & $D_k$ & $\frac{1}{R_p}$ & $\frac{K_{DC}V_{DC,k}^*}{K_\theta}$ \\
			\hline			
		\end{tabular}
		\label{t:tab1}
	\end{center}
\end{table}

\section{Network Dynamics of Grid-Forming Controls} \label{s:Network}
Having established the equivalence among the nodal dynamics of the grid-forming controllers under consideration, this section addresses the network dynamics of the graph with power converters on its nodes and transmission lines across its edges. Transmission line AC power flows are summarized in the following equation \eqref{eq:ACPowerFlow}:

\begin{equation}
\begin{split}
P_k &= \sum_{l=1}^{n} ||V_k||\cdot||V_l||\cdot\big[G_{kl} cos(\theta_k-\theta_l) + B_{kl} sin(\theta_k-\theta_l)\big] \\
Q_k &= \sum_{l=1}^{n} ||V_k||\cdot||V_l||\cdot\big[G_{kl} sin(\theta_k-\theta_l) + B_{kl} cos(\theta_k-\theta_l)\big]
\end{split}
\label{eq:ACPowerFlow}
\end{equation}

Dealing with these highly nonlinear functions of nodal angle differences poses a significant challenge. To simplify the theoretical analysis, AC power flow is transformed to so called DC power flow by linearizing the AC equations around the nominal trivial operating point $P_k^* = P_k = Q_k^* = Q_k =0, \forall k$. Resulting linearized power flows are as follows:

\begin{equation}
\begin{split}
P_k &= \sum_{l=1}^{n} B_{kl} \big(\theta_k - \theta_l\big)\\
Q_k &= \sum_{l=1}^{n} B_{kl} \big(V_{m,k} - V_{m,l}\big)
\end{split}
\label{eq:DCPowerFlow}
\end{equation}

Further simplifying assumption is that all converters are tuned identically, meaning the same $M_k = m, D_k = d, V_{m,k}^* = V_m^*$ for $\forall k$. Through linearization and vectorization of nodal equation \eqref{eq:VSMfinal} for all nodes $k \in [1,..,n]$ one arrives at linearized VSM network dynamics in dq-frame \eqref{eq:VSMnetwork}:

\begin{equation}
\begin{split}
\frac{d}{dt} \theta_\delta &= \omega_\delta \in \mathbb{R}^{n\times1}\\
\frac{d}{dt} \omega_\delta &= -\frac{1}{m} L_B \theta_\delta -\frac{d}{m} \omega_\delta \in \mathbb{R}^{n\times1}\\
\frac{d}{dt} V_{m\delta} &= -\frac{1}{\tau_f} \big[L_B R_q + I_n\big] V_{m\delta} \in \mathbb{R}^{n\times1}
\end{split}
\label{eq:VSMnetwork}
\end{equation}

Representing the first two dynamics as a system \eqref{eq:VSMnetworkfinal}:

\begin{equation}
\frac{d}{dt}\begin{bmatrix}
\theta_\delta\\
\omega_\delta
\end{bmatrix}
=
\underbrace{\begin{bmatrix}
0_{n \times n} & I_{n}\\
-\frac{L_{B}}{m} & -\frac{d}{m}I_{n}
\end{bmatrix}}_\text{$\mathcal{L} \in \mathbb{R}^{2n \times 2n}$}
\begin{bmatrix}
\theta_\delta\\
\omega_\delta
\end{bmatrix} \in \mathbb{R}^{2n \times 1}
\label{eq:VSMnetworkfinal}
\end{equation}

By $L_B \in \mathbb{R}^{n\times1}$ one denotes the Laplacian matrix of susceptances $B$ of the network (graph). By $m$ and $d$ one denotes inertia and damping constants of any single power converter an any node. Using the equivalence from the last section, subsequent discussion is valid for all three control techniques under consideration depending on the values of $m$ and $d$ selected.

Analyzing the system \eqref{eq:VSMnetworkfinal} one can claim the following:

\begin{itemize}
\item Laplacian matrix $L_B$ has $n$ eigenvalues $\lambda(L)$. One of them is strictly at 0, others are real and have strictly positive real parts. The more the connectivity of the corresponding graph of $L_B$ the greater the non-zero eigenvalues are.
\item Eigenvalues of the so called "larger" Laplacian matrix $\mathcal{L}$ depend on eigenvalues $\lambda(L)$ as follows \eqref{eq:Eigenvalues}:

\begin{equation}
\eta = -\frac{d}{2m} \pm \frac{1}{2} \sqrt{\frac{d^2}{m^2} - \frac{4\lambda}{m}}
\label{eq:Eigenvalues}
\end{equation}

So there is one eigenvalue at $\eta = 0$, one at $\eta = -d/m$ and $2n-2$ potentially complex eigenvalues with strictly negative real parts.
\item Using appropriate matrix similarity transformation $T \in \mathbb{R}^{2n \times 2n}$, the dynamics in \eqref{eq:VSMnetworkfinal} can be expressed in the form of average angle/speed and respective difference vectors $[\theta_{AVG};\omega_{AVG};\Delta \theta_\delta; \Delta \omega_\delta]$. Next, it can be shown that $\eta = 0$ eigenvalue corresponds to the $d\theta_{AVG}/dt = \omega_{AVG}$ dynamics. The remaining state dynamics do not depend on $\theta_{AVG}, \omega_{AVG}$, so the corresponding $L' \in \mathbb{R}^{2n-2 \times 2n-2}$ is Hurwitz, implying asymptotic stability of the states $[\Delta \theta_\delta; \Delta \omega_\delta]$ around $[0;0]$
\item It may be shown that by adding the power disturbance of the form $P_d = [P_{d,AVG}; \Delta P_d]$ angle difference vector $\Delta \theta_\delta$ and average speed $\omega_{AVG}$ converge to some non-zero disturbance-dependent constants. Average angle $\theta_{AVG}$ continues integrating as a result of non-zero $\omega_{AVG}$ and corresponding eigenvalue $\eta = 0$.
\item Second largest eigenvalue of $\mathcal{L}$, $\eta_2$ determines the rate of convergence of the states to steady state. From \eqref{eq:Eigenvalues} the smaller (more negative) the $\eta_2$ the more connected the graph, the faster the convergence and potentially the more oscillatory the transient. Indeed, smaller $\eta_2$ implies greater $\lambda$'s - connected graph. Also smaller $\eta_2$ means transient exponential decaying at faster rate. Finally, greater $\lambda$'s can lead to imaginary components in $\eta$'s (negative values under square root), resulting in oscillatory (underdamped) transient decay of the states.
\end{itemize}

From these it can be deduced that in order to have the desired network-level dynamics (at least within the idealized setting at hand), one needs to have a large $d/m$ ratio, such that the eigenvalues $\eta$'s are far into the left half-plane and strictly real even for highly connected graph (high $\lambda$'s). This, in turn, means high damping constant $d$ and low inertia constant $m$, resembling droop parameters.

On the other hand, the ROCOF has to be kept within specified standards to comply with practical limits of protection devices. Thus inertia parameters $m$ cannot be too low. Overall, the desired parameter combination is low enough $m$ (to limit ROCOF) and a high $d$ (to stabilize and accelerate the transient even for highly connected networks). Table \ref{t:tab1} can be used for translating these requirements to the gains of respective controllers.

Last but not least, despite the fact that equivalence was established among the three grid-forming controllers at hand, there are practical advantages associated with matching control that make it an attractive option for a further research. First, matching control does not assume a constant stiff DC source behind the inverter. Second, instead of controlling active power directly, matching control regulates the DC voltage and attaches $\omega$ to it. As it was observed, the net result is the same nodal grid-forming ($\theta,\omega,V$-forming) operation of a power converter. However, matching does a step in the direction of right causality, assuming that some power source behind the DC-link is the main generator of energy, not the inverter itself as in case of VSM or droop control.

\section{Conclusion and Outlook} \label{s:Conclusion}
To sum up, this work addressed the equivalence of three grid-forming control techniques (VSM, droop and matching) using theoretical analysis of a simplified nodal converter and network system. For nodal analysis the AC voltage/current controllers are assumed to be ideal. Also in case of VSM/droop the DC voltage was assumed to be stiff. For network analysis, all node converters are assumed to be identical and the AC power flow was linearized around trivial operating point. These simplifications allow the theoretical study of the grid-forming controllers at hand. Through this analysis the equivalence of the techniques was shown both in nodal and network-level dynamics. Droop and matching counterparts of the VSM dynamic parameters were presented. Network-level analysis showed how the VSM-equivalent parameters should be selected to achieve network-level objectives of a stable, fast and non-oscillatory transients while respecting the nodal constraints, such as rate of change of frequency (ROCOF). Overall, with established equivalence of the techniques, one would suggest to use matching control as it incorporates the DC voltage dynamics and makes a step towards a right power flow causality.

As an outlook, one foresees the work in the direction of matching control. Establishing fully correct causality between grid and RES, current limiting during transients, network planning issues - to name just a few challenges yet to be addressed.

\balance


\begin{thebibliography}{00}
\bibitem{Kroposki} B. Kroposki et al., "Achieving a 100\% Renewable Grid: Operating Electric Power Systems with Extremely High Levels of Variable Renewable Energy," in IEEE Power and Energy Magazine, vol. 15, no. 2, pp. 61-73, March-April 2017, doi: 10.1109/MPE.2016.2637122.
\bibitem{Peng} Q. Peng, Q. Jiang, Y. Yang, T. Liu, H. Wang and F. Blaabjerg, "On the Stability of Power Electronics-Dominated Systems: Challenges and Potential Solutions," in IEEE Transactions on Industry Applications, vol. 55, no. 6, pp. 7657-7670, Nov.-Dec. 2019, doi: 10.1109/TIA.2019.2936788.
\bibitem{Rocabert} J. Rocabert, A. Luna, F. Blaabjerg and P. Rodríguez, "Control of Power Converters in AC Microgrids," in IEEE Transactions on Power Electronics, vol. 27, no. 11, pp. 4734-4749, Nov. 2012, doi: 10.1109/TPEL.2012.2199334.
\bibitem{Poolla} B. K. Poolla, D. Groß and F. Dörfler, "Placement and Implementation of Grid-Forming and Grid-Following Virtual Inertia and Fast Frequency Response," in IEEE Transactions on Power Systems, vol. 34, no. 4, pp. 3035-3046, July 2019, doi: 10.1109/TPWRS.2019.2892290.
\bibitem{Tayyebi} A. Tayyebi, D. Groß, A. Anta, F. Kupzog and F. Dörfler, "Frequency Stability of Synchronous Machines and Grid-Forming Power Converters," in IEEE Journal of Emerging and Selected Topics in Power Electronics, vol. 8, no. 2, pp. 1004-1018, June 2020, doi: 10.1109/JESTPE.2020.2966524.
\bibitem{DArco} S. D'Arco and J. A. Suul, "Equivalence of Virtual Synchronous Machines and Frequency-Droops for Converter-Based MicroGrids," in IEEE Transactions on Smart Grid, vol. 5, no. 1, pp. 394-395, Jan. 2014, doi: 10.1109/TSG.2013.2288000.
\bibitem{Curi} S. Curi, D. Groß and F. Dörfler, "Control of low-inertia power grids: A model reduction approach," 2017 IEEE 56th Annual Conference on Decision and Control (CDC), Melbourne, VIC, 2017, pp. 5708-5713, doi: 10.1109/CDC.2017.8264521.
\bibitem{Arghir} C. Arghir and F. Dörfler, "The Electronic Realization of Synchronous Machines: Model Matching, Angle Tracking, and Energy Shaping Techniques," in IEEE Transactions on Power Electronics, vol. 35, no. 4, pp. 4398-4410, April 2020, doi: 10.1109/TPEL.2019.2939710.
\end{thebibliography}
\end{document}